\documentstyle[psfig,epsf,referee]{mn}

\newcommand{\bed}{\begin{displaymath}}
\newcommand{\eed}{\end{displaymath}}
\newcommand{\beq}{\begin{equation}}
\newcommand{\eeq}{\end{equation}}
\newcommand{\beqa}{\begin{eqnarray}}
\newcommand{\eeqa}{\end{eqnarray}}

\newcommand{\ie}{{\it i.e.}}
\newcommand{\etal}{{\it et al.}}          

\newcommand{\acta}[3]{#3,        Acta Astron. {\bf #1}\rm, #2.}

\newcommand{\astrzh}[3]{#3,      Astron. Zh. {\bf #1}\rm, #2}
\newcommand{\nat}[3]{#3,         Nature {\bf #1}\rm, #2}

\newcommand{\apj}[3]{#3,         ApJ  {\bf #1}\rm, #2}
\newcommand{\aj}[3]{#3,          AJ  {\bf #1}\rm, #2}

\newcommand{\aeta}[3]{#3,        A\&A  {\bf #1}\rm, #2}
\newcommand{\aetas}[3]{#3,       A\&A Suppl.  {\bf #1}\rm, #2}

\newcommand{\mnras}[3]{#3,       MNRAS {\bf #1}\rm, #2}

\newcommand{\physrev}[3]{#3,     Phys. Rev. {\bf #1}\rm, #2}

\newcommand{\pasp}[3]{#3,        PASP  {\bf #1}\rm, #2}

\psrotatefirst

\title[Microlensing signatures]{Chromatic and spectroscopic signatures of
          microlensing events as a tool for the 
          gravitational imaging of stars}
\author[D. Valls--Gabaud]{David  Valls--Gabaud$^{1,2}$ \thanks{Email: {\tt dvg@astro.u-strasbg.fr ; dvg@ast.cam.ac.uk} } \\ 
   $^1$ URA CNRS 1280, Observatoire de Strasbourg,
               11, rue de l'Universit\'e, 
               67000 Strasbourg, France \\
   $^2$ Royal Greenwich Observatory, Madingley Road,
 	       Cambridge CB3 0EZ, UK.
             }

\date{\raisebox{10pt}[0pt]{ }}

\pagerange{\pageref{firstpage}--\pageref{lastpage}}

\pubyear{1997}

\begin{document}

\label{firstpage}

\maketitle           
 
   \begin{abstract}
The detection of  microlensing events from stars
in the Large Magellanic Cloud  and in the Galactic bulge  raise 
important constraints on
the distribution of dark matter and on galactic structure, although
some events may be due to a new type of intrinsic variability.
When lenses are relatively close to the sources,  
we predict that chromatic and spectroscopic effects are likely to appear 
for a significant fraction of the microlensing events. These effects are
 due to the differential amplification of 
the limb and the centre of the stellar disc, and
present a systematic dependence with 
wavelength and time that provide an unambiguous signature of a microlensing 
event (as opposed to a new type of intrinsic stellar variability). 
In addition, their measure would provide a direct constraint on stellar atmospheres,
allowing a 3-dimensional reconstruction or imaging of its structure, a
unique tool to test the current models of stellar atmospheres.  
\end{abstract}


\section{Introduction}

 Following the pioneering idea by Paczi\'nsky (1986), the  detections of several microlensing events from  stars
in the Large Magellanic Cloud (Alcock \etal\ 1993, 1996; Aubourg \etal\ 1993) and of more than 100 from stars in the Galactic bulge 
(Udalski \etal\ 1993, 1994; Alcock \etal\ 1995, 1997) have raised 
 important constraints on the structure of the different components
 of our Galaxy. The associated databases containing millions of light
curves are also a goldmine for studies in stellar variability (see Ferlet \etal\ 1997, and references therein). 

Yet doubts may reasonably be cast on the detections of at least some microlensing events. The
two EROS candidates are variable stars : EROS 1 is a Be
star (Beaulieu \etal\ 1995), while EROS 2 is an eclipsing binary 
(Ansari \etal\ 1995). Although 
spectra of other candidates taken after the event (Della Valle 1994) or during the event itself
(Benetti \etal\ 1995; Lennon \etal\ 1996) show no apparent signs of anomalies, the possibility
remains that these events are just detections of a new type of variable
star. The detection of ``bumpers'' (Alcock \etal\ 1996), a hitherto unknown type
of variable star, and the possibility that dwarf novae and cataclismic variables mimick some events
(Della Valle \& Livio 1996) are particularly worrying, even though the distribution  of the events in the HR diagram indicates that not all the events can be explained
by such phenomena.
 In addition to continuous and follow-up monitoring, it seems important to find
unambiguous proofs that the events detected can indeed be associated with
microlensing events.

In this context it is interesting to note that 
the large rate observed towards the bulge may be accounted
for if about 50 to 90\% of the events are due to lenses within the bulge 
itself (Kiraga \& Paczy\'nski 1994), 
particularly if the bar is 
oriented close to the line of sight. 
The same applies if many of the lenses are within the LMC (Wu 1994; Sahu 1994)
 or in the halo and bulge of M31.

If this is the case, a significant number of 
microlensing events should  
 present characteristic chromatic (Valls-Gabaud 1994; Witt 1995; Bogdanov 
\& Cherepaschuk 1995;  Gould \& Welch 1996)
 and spectroscopic signatures  (Valls-Gabaud 1994; Loeb \& Sasselov 1995), and
we present in the {\it Letter} detailed predictions for both effects.  
These 
effects arise from the differential amplification between the limb and
the inner regions of the stellar disc, where the emerging intensities 
are different due to the temperature gradient in the atmosphere.  The
differential amplification also produces a non-zero
degree of polarisation during the events (Simmons \etal\ 1995ab), and a
spectroscopic line shift (Maoz \& Gould 1994).

\section{Microlensing of stars as extended sources}

The amplification produced by a lens of mass $M_L$ at a distance
$D_{\rm OL}$ on a source assumed to be point-like at a distance $D_{\rm OS}$  (Refsdal 1964; Liebes 1964) is $A_p(y) \; = \; (y^2+2) ( y \sqrt{y^2+4} )^{-1} $ 
where $y$ is the dimensionless distance between the projected position
of the lens and the centre of the source,
$ y  =  \sqrt{ \alpha_o^2  +  (t-t_o)^2/\tau^2 } $
and
$ \alpha_o = \theta_o  / \theta_{\rm E} \; , \; \tau  =  D_{\rm OL} \theta_{\rm E} v_{\perp}^{-1} $
are respectively the minimum dimensionless impact parameter at time $t_o$,  and time scale of the event, with $v_{\perp}$ the transverse velocity of the lens.
$\theta_{E}$  is
the angular Einstein radius 
$
\theta_{\rm E}   =   \left( 4 G M_L D_{\rm LS} / c^2 D_{\rm OL} D_{\rm OS}\right)^{1/2}  
$
where $D_{\rm LS}$ is the distance separating the lens and the source.
Since all scales involved are much larger than the wavelength of light, the
geometrical optics approximation is valid, and the events should not show 
diffraction, interference or chromatic effects. 
Yet, given the colour-magnitude diagrams of the bulge and the LMC, and the
sensibility limits of the present surveys, one expects that between 
one and two thirds of the events are associated with giant stars, so that the point
source approximation  might be questionable. 
If 
the angular radius $\theta_*$ of a star of radius $R_*$ seen at a 
distance 
 $D_{\rm OS}$ is comparable with the angular Einstein radius $\theta_{E}$ of the lens,  
then effects arising from the finite size of the stars will become apparent.
Thus it is only when either the lens mass $M_L$ or the relative distance 
$D_{\rm LS} / D_{\rm OS}$  is  small (or both) that these angles are
similar.  Although the conditions for this to be true might seem
contrived, there will always be a number of microlensing
events where the lens will transit through the stellar disc (Gould 1994;
Witt 1995). 
The fraction ${\cal F}$ of all lensing events which have a maximum impact parameter smaller than
the Einstein radius is simply
${\cal F}  = \ <\theta_*> < 1 / \theta_{\rm E}>$
, where average must be taken over the distribution of distances of both
lenses and sources, and hence depends  on the galactic model adopted.
Since the fraction increases like $M_L^{-1/2}$, 
if the lenses are sub-stellar objects the fraction may reach unity for a
large range of $D_{\rm LS}$ distances.
However for lenses of 0.1 M$_\odot$ and  typical fractional distances 
$D_{\rm LS}/D_{\rm OS}$  from 0.1 (bulge) to about 
5 10$^{-3}$ (M31), the fraction of transit events goes from 31\% to 16\% for
typical clump giants of 10 $R_\odot$. Since they constitute at least a third of the
sources, the {\it minimum} fraction of transit events is around 10\% for the
bulge and about 5\% in M31, for  0.1 M$_\odot$ lenses. For these events, the
sizes of the stars has to be taken into account.

\begin{figure}
\psfig{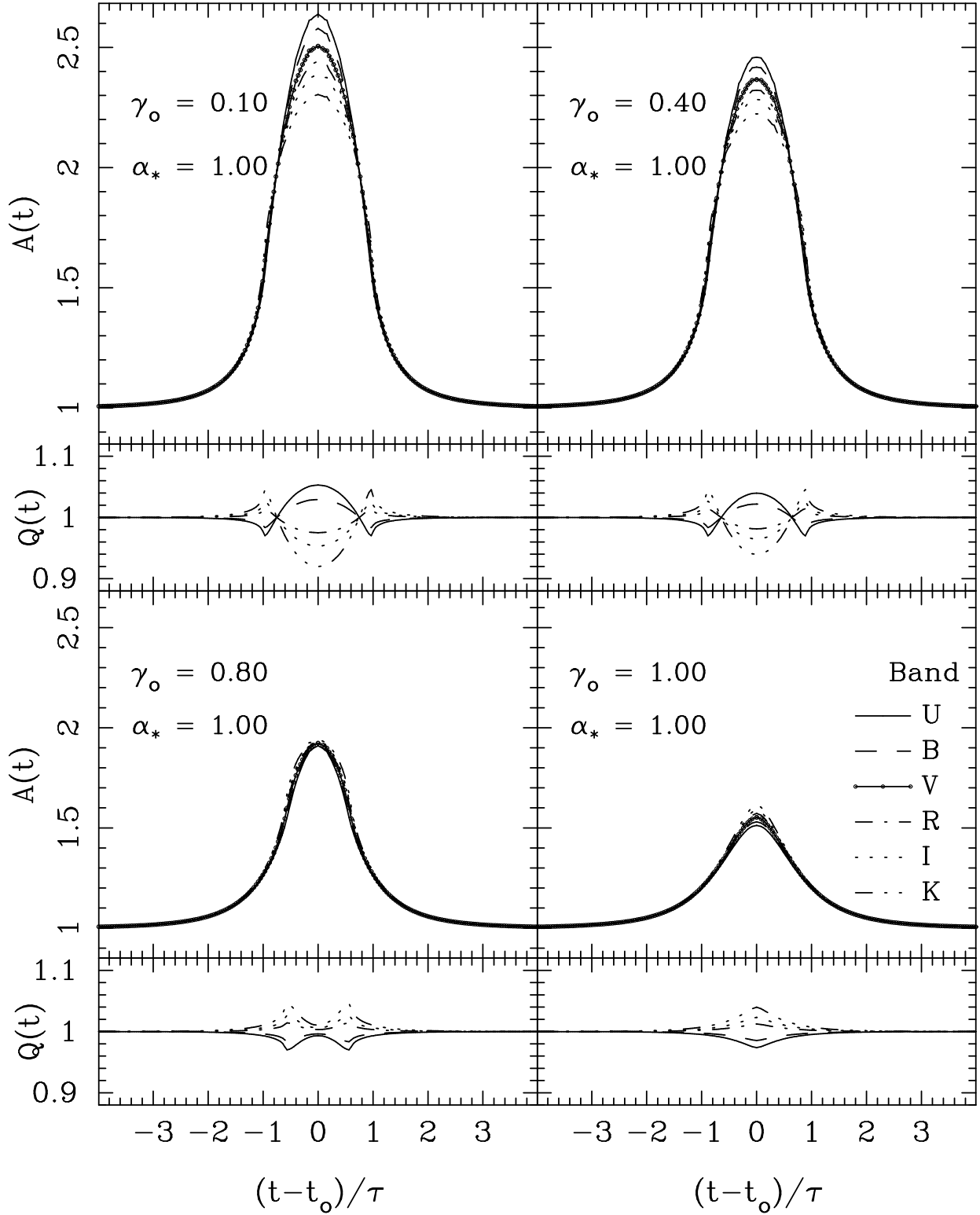} 
\caption{Chromatic signatures of the microlensing of a giant red star of $T_{\rm eff}$ = 4000 K, 
$\log g$=2.0 and solar metallicity, for the indicated geometrical
parameters [$\gamma_o = \gamma(t_o)$]. The magnifications for the  extended Johnson photometric 
broad bands 
are calculated assuming a square root non-linear limb darkening profile. 
The lower panels give the relative magnification relative to the one in the
 Johnson $V$ band, 
$Q = A_\lambda / A_V$ which reflects the relative change in colour index $ \Delta(m_\lambda - m_V) \approx 1.086 (1 - Q) $. }
\label{fig:chromatic}
\end{figure}

\section{Chromatic and spectroscopic signatures}

The inclusion of the
finite size of the sources was first analysed by Bontz (1979) and 
more recently by Nemiroff \& Wickramasinghe 
(1994), Gould (1994) and Witt \& Mao (1994) for
the idealised case of a uniform brightness source, but 
little attention has been paid so far to the amplification produced by
actual stars. Yet it is well known that even in the Eddington approximation
the difference in intensity between the centre and the limb of a star reaches
almost a magnitude.
In the general case then, the magnification $A_e$ of an extended source with  
specific intensity 
 ${\cal I}(\varpi)$ is given by the integration over the extension
${\bf \Omega}$ of the source of the amplification produced on an infinitesimal
area on the source at an angular distance $\theta$ from the point lens 
$
A_e(\theta) \; = \; \int_{\bf \Omega} \; d\varpi \; {\cal I}(\varpi) \; A_p(\theta) 
\; / \;  \int_{\bf \Omega} \; d\varpi \; {\cal I}(\varpi)
$.
Assuming that the star is spherically symmetric (\ie, neglecting rotational
flattening, spots, etc.) and using polar coordinates in the source plane and the notation 
$p^2  =  \gamma^2  +  x^2  +  2 \, \gamma \, x  \cos \phi $
the actual magnification of the star becomes
\beq
A_e(\gamma) = \frac{1}{\pi R_\lambda} \; \int_o^{2\pi} d\phi \int_o^1
dx \, x \, {\cal B}_\lambda(x) \, A_p(p\alpha_*)    \label{eqn:mag}
\eeq
where $x=r/R_*$, $\alpha_*=\theta_*/\theta_{\rm E}$, $\gamma=\theta_{\rm LOS}/\theta_*$ is the angular separation between the lens and the centre
of the star in units of the stellar angular radius,  ${\cal B}_\lambda(x) = {\cal I}_\lambda(x) / {\cal I}_\lambda(0)$ is the normalised brightness
profile, and  $R_\lambda = 2 \int dx \, x \, 
{\cal B}_\lambda$ is the dimensionless effective radius at that wavelength. In the case where  ${\cal B}_\lambda(x)$ is
constant, the maximum amplification reduces to $A_{\rm uniform}^{\rm max} =
\left( 1 + 4/\alpha_*^2 \right)^{1/2}$ (Refsdal 1964, Bontz 1979), while for $\gamma\gg$ 1, $A_e \rightarrow A_p$, the point-source limit.
The brightness profile ${\cal B}_\lambda$  is given by the solution of the
radiative transfer equation :
\beq
{\cal B}_\lambda(x) = \int_0^\infty \, \frac{S_\lambda(\tau_\lambda)}{{\cal I}_\lambda(0)} \,\exp\left[-\frac{\tau_\lambda}{(1-x^2)^{1/2}}\right] \, \frac{d\tau_\lambda}{(1-x^2)^{1/2}} \label{eqn:source}
\eeq
where ${\cal I}_\lambda(0)$ is the emerging intensity at the centre of the
star and $S_\lambda$ the source function. 
It is the explicit dependence with wavelength of the brightness profile ${\cal B}_\lambda$ and the differential amplification that produce both chromatic and
spectroscopic effects (Valls--Gabaud 1994),  that reflect the temperature
gradient $T(\tau_\lambda)$ in the stellar atmosphere.

Given a stellar atmosphere model, one
derives the corresponding limb profiles for both the continuum and the lines.
We have used the large grid of LTE models from Kurucz (1994) and fitted, for
the both continuum and a selected set of lines, simple analytical expressions. The
reason for this is twofold : first the computational speed is substantially
increased, and second the information content of the limb darkening is limited
 and prevents the use of very detailed models (B\"ohm 1961). 
Since the pioneering studies of Milne (1922), 
the  limb profiles for the continuum are approximated by a linear law
${\cal B}_{\lambda}^{\rm LIN}(x) = 1 - x_{\lambda} \left( 1 - \mu \right)$,
where $\mu = (1-x^2)^{1/2}$. In the modelling of  
stellar atmospheres, however, non-linear laws
have yielded significantly better fits and in particular  
the logarithmic law 
${\cal B}_{\lambda}^{\rm LOG}(x)  = 
  1  -  x_{\lambda}  \left( 1  -  \mu \right)  - y_{\lambda}  \mu  \ln \mu $
and the square root law (D\'{\i}az-Cordov\'es \etal\ 1995) 
 ${\cal B}_{\lambda}^{\rm SQR}(x)  = 
 1  - x_{\lambda}  \left( 1  -  \mu \right)  -  y_{\lambda}  \left( 1  -  \sqrt{\mu} \right)$
seem to provide good fits to most of the computed profiles (Van Hamme 1993), although we stress that 
actual precise data is only available for the Sun. 
In order to predict the different light curves in 
the photometric passbands used in  microlensing monitoring projects, 
the limb darkening coefficients are integrated over these bands, and
our results agree with those published by   Claret \& Gim\'enez (1992) and
Van Hamme (1993), for both the Johnson and Str\"omgren systems.

The important point to note is that since the limb darkening coefficients obviously
depend on wavelength due to the variation in opacity, different limb profiles
will appear depending on the photometric band.  This strong
dependence with wavelength implies that the same microlensing event will present
quite different light curves in different photometric bands, as shown in 
Fig. \ref{fig:chromatic}. 

 We can understand the behaviour of the light curves as a function of
wavelength  by noting
that  the effective radius
$R_\lambda$ increases from 0.64 in the $U$ band to 0.90 in
the $K$ band for the typical red giant star in Fig. \ref{fig:chromatic}. Thus, the longer the wavelength,
the more uniform the star is, while in the near UV stars are more
sharply peaked, giving a larger amplification when the impact
parameter is small. At large impact parameters the strong decrease in
brightness at shorter wavelengths takes over and the magnification becomes
smaller, relative to larger wavelengths. Using the actual limb darkening
profiles, the linear Milne law or the non-linear profiles do not make any
significant difference ($\Delta A/A \leq 0.01$) in the final magnifications, and hence the use of the
linear approximation is fully justified. 

One should stress that all these predictions are based on theoretical models
which have never been tested in such detail. There may well be further
complications, such as for example, at larger wavelengths, in the far IR or sub-mm range, where the limb can actually be
brighter (at least in the Sun). In extended
envelopes, the increased opacity in the UV can produce a larger $R_\lambda$ in
this domain, and decrease the contrast $Q$, while in the
X-ray domain, the coronal emission is very patchy and a smooth, spherically
symmetric distribution is meaningless. There may
be also complications related to the presence of chromospheres, which change
the temperature gradient and hence the limb profiles. The spectral lines
will be more sensitive to this effet (see below).
To first order, the chromatic effect does not depend on the colour of the stellar source, since the integrated intensity of the star does not appear
in Eq. \ref{eqn:mag}. However, there will be some dependence since limb darkening
profiles do depend on the spectral type of the star, as well as on the
luminosity class. In general, the limb darkening increases when 
the gravity or temperature decreases and also when the metallicity increases,
so the giants of the Galactic and M31 bulges are ideal probes. 

\begin{figure}
\psfig{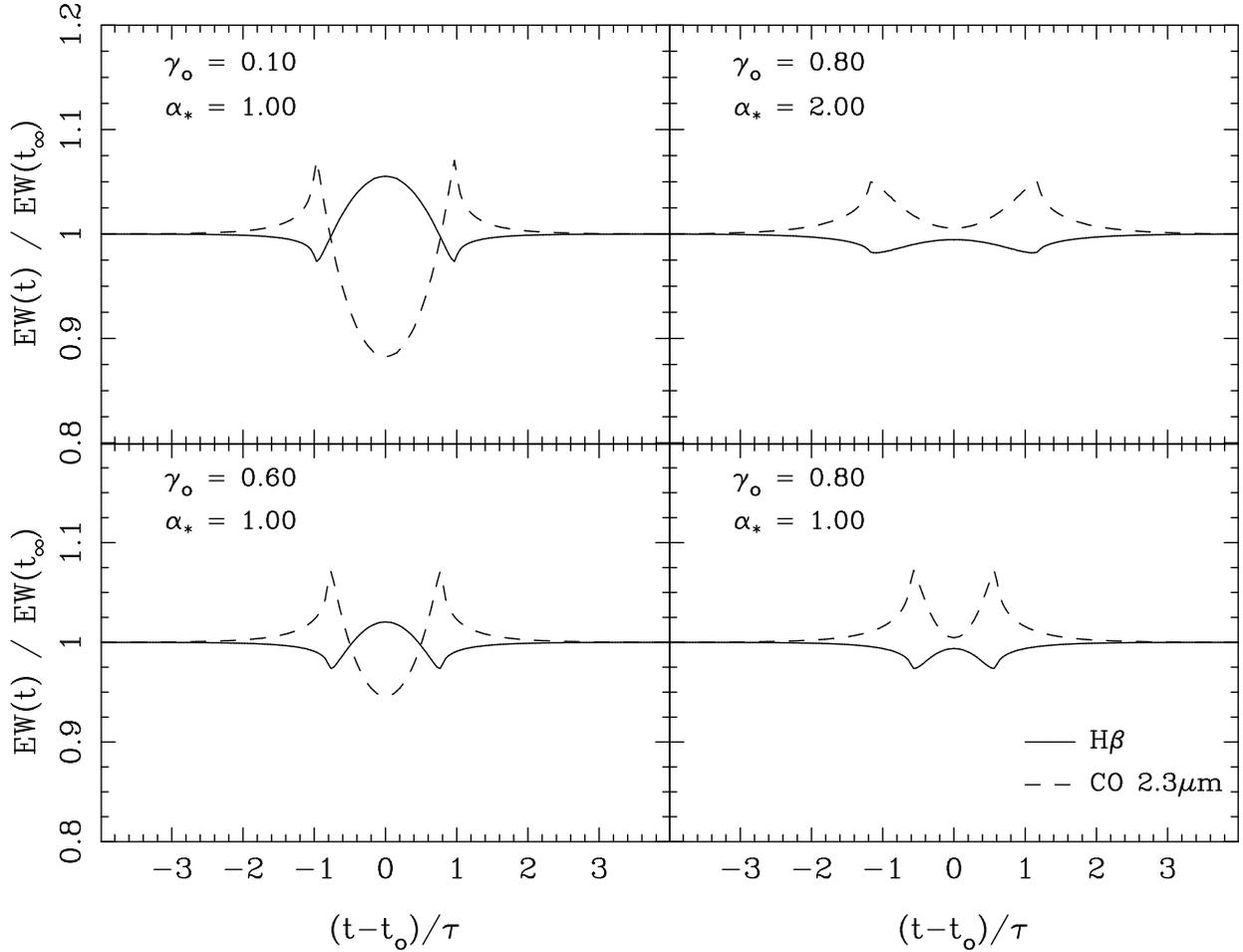}
\caption{Spectroscopic signatures of microlensing, for the same giant star
considered in Fig. \ref{fig:chromatic}. and two typical lines : H$\beta$ and CO 2.3 $\mu$m. On
the right panels the effect of increasing the stellar diameter (from $\alpha_*=1$  to $\alpha_*=2$) is illustrated for the otherwise same transit conditions.}
\label{fig:spec}
\end{figure}

The best way to measure the chromatic effect is to observe the event in
photometric passbands as widely separated in wavelength as possible, ideally
$U$ and $K$ which are the most discriminatory for giant stars. It is unfortunate
that the current surveys use bands as close as  $V$, $R$  and $I$, where the effect
is predicted to be very small. As Fig. \ref{fig:chromatic} shows, the relative magnifications
$Q$ (with respect to the $V$ band) are smaller than 10\%, that is, the change in colour
index $\Delta(m_\lambda - m_V) \approx 1.086 (1 - Q)$  is less than 0.1 magnitudes, hence it is also essential to do {\em differential}, rather than absolute, photometry to measure the effect. 
 Present technology may achieve 0.001 mag (Young \etal\ 1991), although
a precision of 0.01 will be enough. Finally,
the use of narrow-band Str\"omgren photometry is also recommended, since the
Str\"omgren bands probe different opacity regimes where the source function 
changes strongly. The wavelength and time dependence of the light curves 
are unlike those produced by any intrinsic stellar variability, and hence constitute a unique 
signature of microlensing events.  

The spectroscopic signature arises, once again, because 
line profiles change significantly across the disc, although in this case the
variation will depend on the mechanism of
formation of the line. For example strong resonant lines, like Ly-$\alpha$ or Ca {\sc ii} H+K, Mg {\sc ii} $h$+$k$ present a decreasing  relative depression with respect to the continuum, and hence appear brighter at the limb, while 
pure absorption lines disappear at the limb. 
Lines are far more sensitive to the actual gradient of temperature in
the atmosphere, since they probe a larger range of depths, while others
are more sensitive to pressure or turbulence. However, 
since the observations of the variability of the line profiles seems difficult with
the present technology, a more appropriate observable is the equivalent
width (EW) of the line. For very strong lines, narrow-band photometry
in the appropriate range could also be used (Loeb \& Sasselov 1995), but
the availability of real-time spectra makes this less practical.
 To illustrate the predicted behaviour of the EWs 
during a microlensing event\footnote{Detailed predictions for any type of star of metallicity between 0.002 and 0.04 in the wavelength range 1200\AA--5$\mu$m are
available upon request.}, Fig. \ref{fig:spec} presents the predictions for the Balmer H$\beta$ and CO 2.3 $\mu$m lines, the
CO forming at lower temperature, much higher in the atmosphere, and is
limb-brightened. It is
immediately apparent that the amplitude of the spectroscopic effect may reach
20\%. In the case of  H$\beta$, which is limb-darkened, the EW first
decreases when most of the flux originates from the limb, then increases 
when the lens is probing the central areas of the star. For the vibrational-rotational
bands of CO (from 2.2 to 4.6 $\mu$m), the opposite behaviour is present, and the temperature gradient is probed at higher altitude in the atmosphere. Once again, the temporal variation of the lines during the event constitute a unique
signature of microlensing, unlike any variation produced by intrinsic stellar 
variability. Note also that the time scale
is no longer the Einstein ring crossing time, but rather the stellar 
crossing time $\sim R_* v_\perp (1 - D_{\rm LS}/D_{\rm OS})$ (see Fig. 
\ref{fig:spec}). 

\section{Discussion}

The gravitational imaging of stars is then possible, where
not only the brightness distribution on the stellar disc is measured (via
the solution of the integral equation in Eq. \ref{eqn:mag}), but also the source
function can be recovered, solving Eq. \ref{eqn:source}, hence giving a 
3-dimensional picture of the stellar atmosphere. This is unprecedented in
astrophysics, where only the Sun and very few nearby stars are
spatially resolved, either directly (e.g., Gilliland \& Dupree 1996) or
by interferometry (e.g.,  Baldwin \etal\ 1996).    

Actual techniques for the inversion of Eqs. \ref{eqn:mag} and \ref{eqn:source}
cannot fit within the scope of this {\it Letter}, and will be dealt with elsewhere.
We simply note here the important implications of stellar imaging.
The predictions presented here are based on model atmospheres, in LTE or NLTE, and the expected limb darkening
coefficients have seldom been compared with observations others than
the solar ones. The reason is that the only way to
measure these coefficients  (besides the solar case) accurately is using eclipsing binaries, which must have circular orbits, 
well-detached components, and produce total eclipses. Even then, 
 the derived coefficients correlate with the assumed sizes of the
stars (e.g., Tabachnik 1969), so the measure of limb darkening coefficients
in microlensing events is a unique opportunity to test stellar atmosphere
models. Note however that the inclusion of non-linear terms produces insignificant differences in the resulting magnification, yet even the
measure of the linear coefficients is important, for they provide an
estimation of the contribution function of the continuum (basically the
integrand of Eq. \ref{eqn:source}).  This in turn gives a direct estimation
of the temperature profile, something that is known in very few stars
(e.g. Matthews \etal\ 1996) and that provides unprecedented constraints on
the opacity sources. This technique, applied to the lines (ideally the
profiles, but the EWs are still extremely important) may  lead to the element abundance imaging of
the star. In general, atmospheric diagnostics will be possible, by selecting
lines sensitive to pressure, or temperature, etc. 

Unlike Doppler or Zeeman-Doppler imaging, the gravitational imaging is
 unbiased towards bright stars with large spots, although we note that the
presence of spots will considerably complicate the inversion unless the
periods are much longer than the time scale of the event. 
Complications may also
arise from blending (Kamionkowsky 1995) and extinction. 
The ongoing real-time follow-up teams, PLANET (Albrow \etal\ 1997) and 
GMAN (Abe \etal\ 1997), could adapt their
observing strategies to measure these signatures, by increasing the time
sampling, widening the wavelength range covered and obtaining the highest
resolution spectra possible. 
 
In conclusion, the chromatic and spectroscopic signatures of microlensing
events allow us not only to deduce properties of the lenses, but also, and maybe
more importantly, to open an entire new window on stellar atmospheres.

\section*{acknowledgements}
I am grateful to W. Van Hamme and A. Claret for providing their
latest limb darkening calculations for comparison with ours.

\noindent {\small Note: After this paper was completed, Alcock \etal\ (1997, ApJ, submitted, {\tt astro-ph 9702199}) have reported real-time photometry and spectroscopy of
events associated with a M4 giant in the Galactic bulge. Both signatures
may be present, although further analysis is required for confirmation.}

\label{lastpage}

\end{document}